\documentclass[twocolumn,showpacs,preprintnumbers,amsmath,amssymb,prb]{revtex4-1}
\usepackage{graphicx}
\usepackage{dcolumn}
\usepackage{bm}
\usepackage{float}
\usepackage{color}

\begin{document}

\title{Explaining the x-ray nonlinear susceptibility of diamond and silicon
near\\ absorption edges}
\author{B. Barbiellini$^1$, 
Y. Joly$^{2,3}$,
Kenji Tamasaku$^{4}$}
\affiliation{
$^1$Department of Physics, Northeastern University, Boston, Massachusetts
02115, USA\\
$^2$ Universit\'e Grenoble Alpes, Institut NEEL, F-38042 Grenoble, France\\
$^3$ CNRS, Institut NEEL, F-38042 Grenoble, France\\
$^4$ RIKEN SPring-8 Center, 1-1-1 Kouto, Sayo-cho, Sayo-gun, Hyogo 679-5148, Japan\\
}

\date{\today}

\begin{abstract}
We report the observation and the theoretical explanation
of the parametric down-conversion nonlinear susceptibility at the $K$-absorption edge of diamond and at the $L_{23}$-absorption edge of a silicon crystal. Using arguments similar to those invoked to successfully predict resonant inelastic x-ray spectra, we derive an expression for the renormalization term of the non-linear susceptibility at the x-ray edges, which can be evaluated by using first-principles calculations of the atomic scattering factor $f_1$. Our model is shown to reproduce the observed enhancement of the parametric down-conversion at the diamond $K$ and the Si $L_{23}$ edges rather than the suppression previously claimed. 
\end{abstract}

\maketitle

\section{introduction}
The advent of x-ray free electron lasers (XFELs) 
\cite{xfel1,xfel2a,xfel2b} has enabled advances in the study 
of x-ray nonlinear processes \cite{xfel3}, which are similar to  
nonlinear optics investigations and applications, that have 
followed the invention of the laser in 1960. 
Few x-ray nonlinear processes have already been observed with 
conventional x-ray sources. In particular, 
x-ray parametric down-conversion (PDC) is an intriguing 
second order nonlinear process where
an x-ray pump photon 
of energy $E_p$ decays 
spontaneously into two photons, 
the idler and the signal 
with energies $E_i$ and
$E_s$, respectively.
PDC was first discussed theoretically by Freund and Levine \cite{1} 
and was then observed 
experimentally by Eisenberger and McCall \cite{2a,2b} in 1971.
The first observations of the PDC into the extreme 
ultraviolet (EUV) were reported by Danino and Freund \cite{3}
about a decade later.
This effect has been recently used to visualize
the local optical response to EUV radiation with an
atom scale resolution \cite{natphys}.
Further improvement of PDC has recently been proposed by  
Shwartz and collaborators\cite{stanford}.

\begin{figure}
\includegraphics[width=8.0cm]{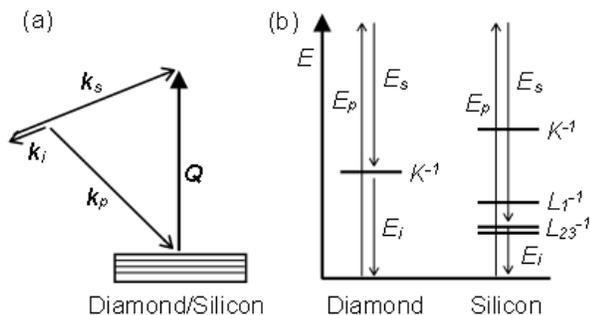}
\caption{Schematics explaining the x-ray PDC:  
(a) momentum conservation law, 
(b) energy conservation laws in diamond and silicon.
The quantities $(E_s,{\bf k}_s)$, $(E_i,{\bf k}_i)$, and 
$(E_p,{\bf k}_p)$ 
are the energy-momenta 
of the signal, idler, and pump 
photons, respectively.}
\label{pdc}
\end{figure}

As illustrated in Fig.~\ref{pdc}, x-ray PDC takes 
place as nonlinear diffraction in 
crystals with energy and momentum conservation laws
given by
\begin{eqnarray}
E_p=E_i+E_s,\\ \nonumber
{\bf k}_p+{\bf Q}={\bf k}_i+{\bf k}_s,
\label{eqpdc}
\end{eqnarray}
where ${\bf k}_s$, ${\bf k}_i$, and ${\bf k}_p$ 
are the wave vectors of the signal, idler, and pump 
photons respectively whereas {\bf Q} is a 
crystal reciprocal
lattice vector.
The momentum conservation is also called the phase-matching condition, 
and rocking the crystal means scanning the phase-matching condition.
Surprisingly, Tamasaku and Ishikawa observed asymmetric rocking curves, 
having not only a peak, but also a distinct dip \cite{4a}. 
If there were no interactions between the PDC
and the Compton scattering, 
the rocking curve should be a Lorentzian peak 
on a smooth Compton background.
Instead, the rocking curve is not
simple but reveals an interference with the 
background Compton process.
The resulting lineshape is similar to 
the one considered by 
Fano \cite{fano}, Vittorini-Oregas and Biaconi \cite{bianconi}.
From the analysis of
the Fano line shape, Tamasaku, Sawada and 
Ishikawa \cite{4b}
extracted the PDC 
nonlinear susceptibility $\chi^{(2)}$.
These authors noticed that a sharp peak 
of the nonlinear susceptibility 
reveals a strong resonant enhancement at the $K$-absorption 
edge of diamond.
This observation contradicts previous models claiming 
that the core resonance suppresses the nonlinear 
process rather than enhances it \cite{5}.

We present here a renormalization
factor $\eta$ that captures the 
behavior of the nonlinear susceptibility 
$\chi^{(2)}$
at the $K$-absorption edge of diamond and at 
the $L_{23}$ absorption edge of Si. 
A similar term has recently been used 
to explain the enhancement of the cross section observed in the
resonant inelastic x-ray scattering (RIXS)\cite{6}.
The factor $\eta$ is derived by using  
the renormalization-group method that describes the variation
in the effective coupling constant under changes of scales 
\cite{landau,bogoliubov,volovik},
and the method of dispersion relations applied 
to the atomic scattering factor.

The remainder of this article is organized as follows.
After the introductory remarks contained in Sec. I, 
Sec. II introduces the model and reviews 
the derivation of the renormalization factor $\eta$.
In Sec. III, we present the methods 
for the calculations of the atomic scattering factor $f_1$ 
used to obtain $\eta$ and for the x-ray PDC experiments. 
The theoretical results are presented and 
compared with the experimental results in Sec. IV.
Important predictions of our model 
and the conclusions are reported in Sec. V.
 
\section{Theoretical Model}
As explained by Freund and Levine \cite{1}, 
the origin of the PDC nonlinearity 
is the Doppler shift, where the induced 
valence charge at the 
idler frequency scatters x-rays 
at a different frequency.
An effective theory for this process is therefore equivalent 
to an inelastic x-ray scattering by valence electrons 
oscillating in the idler field.
In the proximity of an absorption 
threshold, the idler photon generates 
a set of virtual
intermediate states involving a core hole
and a corresponding electron excited in a virtual state which
can be described by an effective dielectric 
function $\epsilon$
experienced by the valence electrons. 
Therefore, the effective coupling constant
can be modified through the dielectric response 
if the idler photon energy is tuned near to the binding energy 
of a core electron level in certain materials.

In general, the dimensionless interaction of quantum electrodynamics (QED)
is not constant but varies with the energy scales. 
In fact, the vacuum of QED can be considered as some kind of polarizable medium, 
where virtual pairs of fermions and anti-fermions screen the electric charge. 
For instance, in the standard model of particle physics, 
the coupling varies because of the crossing of particle production 
thresholds as illustrated by Jegerlehner \cite{jeger}.
Similarly, in materials 
such as FeTe or TiSe$_2$ \cite{6}, 
photons with energies at the proximity of 
the $L_3$ absorption threshold
generate a set of virtual intermediate states 
involving a $2p$ core hole and a corresponding electron excited 
in a $3d$ state, which can be described by an effective dielectric 
function experienced by the valence electrons.
Therefore, we can reduce the dimensionless interaction strength $g$ in a material
by introducing photons with energies just below the threshold of x-ray edges, 
thereby increasing dielectric screening. 
Consequently, the x-ray scattering cross section
$\sigma$ can be significantly enhanced due to an 
increased background polarizability.
The connection between the effective coupling $g$ and 
the length scale $\ell=\sqrt{\sigma}$ 
can be described by a renormalization-group equation, 
\begin{equation}
\beta(g) = \frac{d g}{d\ln(\ell)},
\label{eq:RG}
\end{equation}
where the function $\beta(g)=-3g^2$ was
derived from a thermodynamic argument \cite{8}
involving a pressure
needed to set the scale of $\ell$
for a given coupling $g$.
The solution of Eq.~\ref{eq:RG} implies that
the x-ray scattering cross section $\sigma$
is renormalized by the factor,
\begin{equation}
\eta=\exp[\frac{2}{3 \alpha}(\epsilon_1/\epsilon_0 -1)],
\label{eqeta}
\end{equation}
where 
$\epsilon_0$
is the dielectric constant in vacuum,
$\alpha$ is the fine structure constant 
(i.e. the effective coupling $g$ when $\epsilon=\epsilon_0$)
and $\epsilon_1$ is the real part of an effective
dielectric function $\epsilon$.
Near absorption edges, there are
anomalous dispersions that allow 
x-ray PDC to be enhanced within a narrow energy range, 
since $\epsilon_1/\epsilon_0$  
can be larger from unity in this energy domain.
As already mentioned above, the Compton scattering 
mixes with the PDC and
both processes experience the same renormalization factor $\eta$
because they involve the same valence electrons embedded in 
the effective medium described by $\epsilon$.
Zambianchi \cite{zambianchi} noticed
that $|\chi^{(2)}|^2$ is proportional to the cross section.
Hence, the renormalization of the cross section also 
yields a renormalization for $|\chi^{(2)}|^2$ given by
\begin{equation}
|\chi^{(2)}|^2=\eta |\chi^{(2)}_{NR}|^2,
\end{equation}
where $|\chi^{(2)}_{NR}|^2$ 
represents the non-resonant part of the 
non-linear susceptibility.
The real part of the dielectric function $\epsilon_1/\epsilon_0$ 
contained in the exponent of $\eta$ depends
on the atomic scattering factor $f_1$ 
trough the equation
\begin{equation}
\frac{\epsilon_1}{\epsilon_0}=1-\frac{4\pi r_0\rho}{k^2_i}f_1,
\end{equation}
where $k_i$ is the norm of the idler wave-vector, $r_0$ is the 
classical electron radius and $\rho$ is the number of C or Si atoms 
per unit volume. 
The factor $f_1$ can be written using
the Kramers-Kronig transform \cite{henke} as
\begin{equation}
f_1(E)= Z^*+\frac{2}{\pi}
        \int_{0}^{\infty} 
        \frac{\omega f_2(\omega)}
             {E^2-\omega^2} d\omega,
\label{eqf1}
\end{equation}
where $Z^*$ is the effective atomic charge and $f_2$ 
is the imaginary part of the complex scattering factor. 
Thus, $f_1$ becomes negative when the number of anomalous
electron near the absorption edge [given by the second term 
on the right side of Eq.~\ref{eqf1})] is greater 
than the number of electrons.
Moreover, both $f_1$ and $\epsilon_1$ 
contain the Lorentz oscillator for 
the core electrons appearing also in the treatment
of RIXS and in a fit of $|\chi^{(2)}|$ proposed by
Tamasaku, Sawada, and Ishikawa \cite{4b}. Nevertheless,
one should notice that
in resonant PDC, $\epsilon_1$ depends
of the idler energy while in RIXS, $\epsilon_1$ is a function
of the incident energy of the photons \cite{5}.

\section{Computational and experimental methods}
Henke {\em et al.} \cite{henke}
have provided a complete tabulation of 
values for the atomic scattering factor $f_1$ 
calculated for all the elements $Z = 1-92$ 
in the energy range from 50 eV to 30 keV.
However, their atomic like assumption is clearly 
insufficient in the vicinity of absorption 
edges because of condensed-matter effects \cite{nayak}.
For this reason, we have calculated the atomic scattering factor from the program 
FDMNES (standing for finite difference method near-edge structure) \cite{joly}, 
which is a first-principles, free and open source code. The program directly 
calculates the complex scattering factor $f=f_1+if_2$ without using the 
Kramers-Kronig transform given by Eq.~(\ref{eqf1}).
FDMNES is a real-space program, that does not involve Bloch states 
and Brillouin zone samplings. Therefore, whether, 
one considers a molecule or a periodic system, 
FDMNES builds a cluster around the absorbing atom. The cluster’s 
radius is chosen large enough in order to achieve 
convergence with respect to the accuracy of the calculation of the final states.
The program performs density-functional theory (DFT) 
self-consistent calculations in the ground and the excited states. 
The excited-state calculation at a given edge 
involves the presence of the corresponding core-hole. 
The computed wave functions provide 
the matrix elements for the direct calculation of the complex scattering factors. 
We have used the default exchange-correlation potential of FDMNES, which is within the local density approximation (LDA). 
In fact, the gradient correction to LDA 
does not give a notable difference for C and Si.
The core-level energy is adjusted to the experiment.
Many-body corrections have been implemented 
via a time-dependent DFT (TDDFT) kernel \cite{TDDFT} 
containing the Hartree term (which gives the main effects) 
and an exchange-correlation LDA term.
We will check that the present TDDFT procedure 
in Si is validated by a very good agreement 
with reflectivity experiments.
The lattice constants used in the present calculations were
$3.56$ \AA~ and $5.43$ \AA~ for diamond and Si, respectively.
For diamond, $\rho=1.77 \times 10^{23}$ cm$^{-3}$ whereas 
for Si, $\rho=0.50 \times 10^{23}$ cm$^{-3}$.
After obtaining the self-consistent electronic structure,
$f_1$ was computed at the C $K$ edge and 
at the Si $L_{23}$ edge.

To reveal the resonance effect around the absorption edge, 
we measured $|\chi^{(2)}|$ of silicon near the
$L_{23}$ edge in addition to the previous report
on the diamond for the $K$ edge \cite{4b} with $\bm Q=(2,2,0)$.
The experiments were performed at the 27-m undulator beamline BL19LXU 
at SPring-8 described by Yabasi and co-workers \cite{yabashi01}.
The pump photon energy was set to 9.67 keV, 
which is just above the $K$ edge of Zn. 
A (111) silicon plate
with a thickness of 1.0 mm was used as the nonlinear crystal. 
The nonlinear diffraction was measured with
$\bm Q=(1,1,1)$. The scattering plane was taken on the horizontal plane, 
and the polarization of the pump
x-rays is also horizontal.  
The signal photon energies were selected around the silicon $L$ edge by a bent
crystal analyzer with a Ge 220 reflection. 
The photon energy resolution of the analyzer was measured to
be 2.7 eV. 
In order to separate the signal from the pump beam,
a 25-$\mu$m-thick Zn foil was inserted 
before the analyzer.
The signal photon was monitored by a NaI scintillation counter,
whereas the idler photon was deduced by 
the energy conservation given in Eq.~(\ref{eqpdc}).

\section{Results}
\begin{figure}
\includegraphics[width=8.0cm]{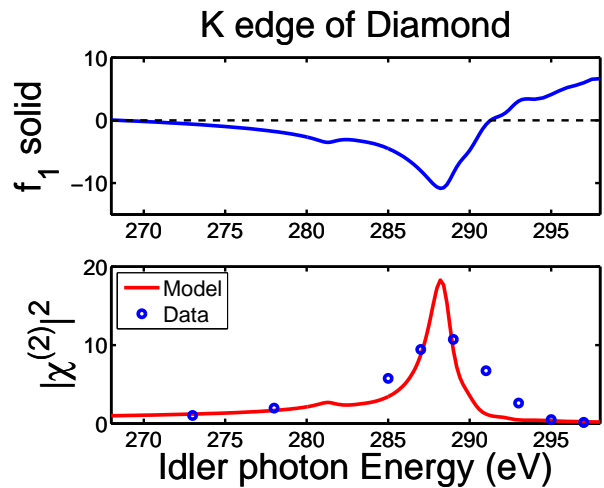}
\caption{(Color online) $K$ edge of diamond:
In the top frame the solid curve represents the model
for $f_1$ as a function of the idler energy.
The label "$f_1$ solid" emphasizes that 
the factor $f_1$ for the solid is different from
the one calculated for free atoms. 
A dashed horizontal line at $0$ highlights 
the $f_1$ changes in sign. 
In the bottom frame the black circles and the 
solid curve are the $|\chi^{(2)}|^2$ 
in units of $|\chi^{(2)}_{NR}|^2$ and the model for 
the renormalization factor $\eta$ 
respectively, plotted against the idler energy.
The experimental data are reproduced 
from the experiment by Tamasaku, 
Sawada, and Ishikawa \cite{4b}.}
\label{f1}
\end{figure}

Figure~\ref{f1} illustrates the resonant behavior of the 
non-linear susceptibility of diamond as 
a function of the idler energy $E_i$.
The $|\chi^{(2)}_{NR}|$ value fitted with our model
is $1.4\times 10^{-16}$ cm$^2$/StC,
where StC means the Gaussian unit of StatCoulomb. 
This value is consistent with perturbation theory  
\cite{foot1}.
The corresponding atomic scattering factor $f_1$ 
calculated within FDMNES is shown 
in the top frame of Fig.~\ref{f1}.
Near the $K$ absorption edge, one can notice the
negative excursion of $f_1$ reaching about $-11$ 
that allows $\epsilon_1/\epsilon_0$ to be larger
from unity near the absorption edge. 
The peak value of the calculated 
$\epsilon_1/\epsilon_0-1$ is 0.033.
A similar dielectric function behavior
can also be seen in the optical data 
from the interstellar dust grains 
of graphite provided by Draine \cite{draine}. 
However, the peak for graphite is smaller 
than the one for diamond 
since the higher density of diamond leads 
$\epsilon_1/\epsilon_0-1$
to be larger by approximately the density ratio.
In the bottom frame of Fig.~\ref{f1}, 
the resonant part of the nonlinear susceptibility
obtained by Tamasaku, Sawada, and Ishikawa \cite{4b} 
is compared to the model.
The overall agreement between the experimental data 
and $\eta$ is good below the threshold energy.
Some discrepancies 
above the absorption threshold can be explained 
by the fact that the present theory neglects some resonant 
fluorescent effects due to the core electrons \cite{6}.  
Interestingly, 
tuning the idler 
photon energy above 295 eV makes 
$\epsilon_1/\epsilon_0$ be smaller
from unity in a certain energy window. 
Thus, $|\chi^{(2)}|^2$ becomes smaller 
than $|\chi^{(2)}_{NR}|^2$ in this region.

\begin{figure}
\includegraphics[width=8.0cm]{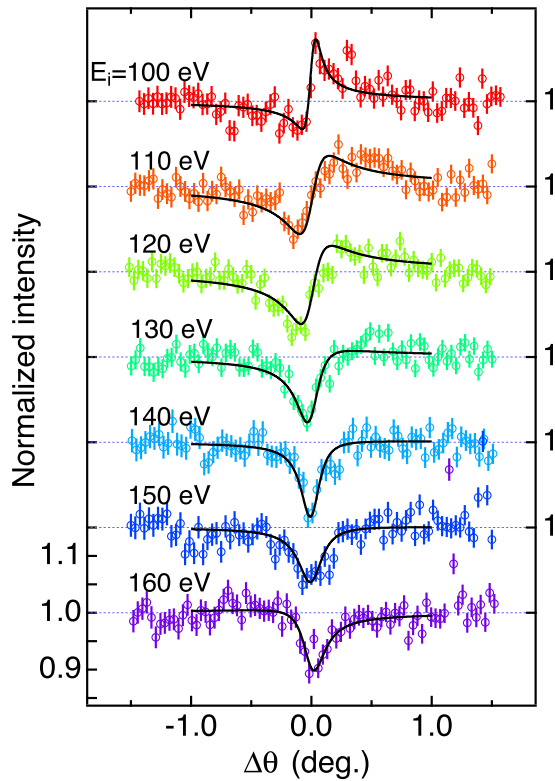}
\caption{(Color online) 
Rocking curves of the 111 nonlinear diffraction of a Si crystal. 
Circles with error bars represent the measurements 
and full lines are fits 
to the data with the Fano formula used by Tamasaku, 
Sawada, and Ishikawa \cite{4b}} 
\label{rock}
\end{figure}

\begin{figure}
\includegraphics[width=8.0cm]{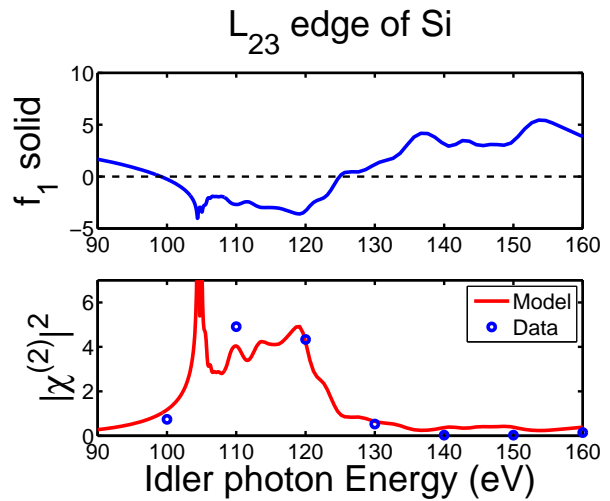}
\caption{(Color online) Same as Fig.~\ref{f1} 
for the Si $L_{23}$ edge.} 
\label{f2}
\end{figure}

The measured Si rocking curves corresponding to
the 111 nonlinear diffraction \cite{SM} are shown 
in Fig~\ref{rock}.
The horizontal axis indicates the deviation angle $\Delta \theta$ 
from the phase-matching condition at each $E_i$.
When $E_i$ approaches the $L_{23}$ 
threshold from below the line shape is asymmetric whereas
when $E_i$ is well above threshold a dip characterizes the spectrum.
The data have been fitted with the Fano 
formula by Tamasaku, Sawada, and Ishikawa \cite{4b}
to extract the ratio $|\chi^{(2)}/\chi^{(2)}_{NR}|^2$ 
shown in the bottom of Fig.~\ref{f2}. 
The top of Fig.~\ref{f2} shows that 
the FDMNES result for $f_1$ near the $L_{23}$ edge
predicts an amplitude of about $-4$ after 
an energy of $E_i=100$ eV. This result
is in good agreement with x-ray 
reflectivity measurements 
by Tripathi {\em et al.} \cite{lodha}
and validates the TDDFT kernel used in FDMNES.
In fact, the peak value of the experimental 
$\epsilon_1/\epsilon_0-1$ is 0.024 
whereas our calculated value gives 0.025.
Clearly, the calculated negative
excursion of $f_1$ leads to good overall
agreement with the experimental ratio $|\chi^{(2)}/\chi^{(2)}_{NR}|^2$.
Thus, the resonant fluorescent effects 
neglected by the theory seem to be small at the $L_{23}$ edge of Si.
We did not measure the pump intensity in the case of Si, therefore 
it was not possible to obtain the non linear susceptibility in units of
cm$^2$/StC.

Interestingly, the enhancement of $|\chi^{(2)}|$ is not 
observed at the Si $L_1$ absorption edge around 150 eV
since $f_1$ does not present any
negative excursion in the $L_1$ edge energy region \cite{nayak}.
Therefore, the sign change in $f_1$ is a crucial condition for the
resonant enhancement of $|\chi^{(2)}|^2$ at a given absorption edge.

\section{Conclusions}
We have refuted the claim 
that the major effect of resonance 
is a decrease in the strength of PDC.
Moreover, we have explained
the factor determining the renormalization of
the nonlinear susceptibility 
$\chi^{(2)}$ at some absorption edges.
In our model,
the square of the nonlinear 
susceptibility $|\chi^{(2)}|^2$ has been connected 
to an effective scattering cross section, 
whose resonant enhancement is a function of 
the scattering factor $f_1$, 
which is strongly modulated when the idler energy 
is varied across resonant absorption edges.
An important prediction of our scheme is that 
the resonant enhancement of $|\chi^{(2)}|^2$ 
for the carbon $K$-edge and the Si $L_{23}$ edge
occurs only when $f_1$ becomes negative otherwise 
a de-enhancement is obtained. This condition for $f_1$ explains 
why the resonant enhancement is not observed at the $L_1$ 
edge of Si.
Our first principles calculations of $f_1$ based on the FDMNES 
are able to faithfully reproduce the dispersion at the edges
needed for the reliability of the model 
since they account for bonding and condensed-matter effects. 
Finally, a related cross-sectional enhancement effect 
has also been observed 
in RIXS experiments \cite{6}. 
Therefore, very accurate RIXS and PDC experiments
by XFELs \cite{foot2} can lead 
to more fundamental theoretical insight.

\begin{acknowledgments}
We are grateful to Grisha Volovik for useful discussions
on the running coupling constants.
This work was supported by the US Department of Energy (DOE), 
Office of Science, Basic Energy Sciences grant number DE-FG02-07ER46352 (core research), 
and benefited from Northeastern University's Advanced Scientific Computation Center (ASCC), 
the NERSC supercomputing center through DOE grant number DE-AC02-05CH11231, 
and support (applications to layered materials) from the DOE EFRC: Center for 
the Computational Design of Functional Layered Materials (CCDM) under DE-SC0012575.

\end{acknowledgments}


\begin{thebibliography}{99}
\bibitem{xfel1}
R. Bonifacio, C. Pellegrini, L.M. Narducci, Opt. Commun. {\bf 50}, 373 (1984).
\bibitem{xfel2a}
P. Emma {\em et al.}, Nature Photonics {\bf 4}, 641 (2010).
\bibitem{xfel2b}
T. Ishikawa {\em et al.}, Nature Photonics {\bf 6}, 540 (2012).
\bibitem{xfel3}
B.W. Adams, {\em Parametric Down Conversion}, 
in {\em Nonlinear Optics, Quantum Optics and Ultrafast Phenomena 
with x-Rays: Physics with x-Ray Free-Electron Lasers}, edited by B.W. Adams 
(Kluwer Academic Publishers, USA, 2003), p.109.
\bibitem{1}
I. Freund and B. F. Levine, Phys. Rev. Lett. {\bf 23}, 854 (1969).
\bibitem{2a}
P. M. Eisenberger and S. L. McCall, Phys. Rev. A {\bf 3}, 1145 (1971).
\bibitem{2b}
P. M. Eisenberger and S. L. McCall, Phys. Rev. Lett. {\bf 26}, 684 (1971).
\bibitem{3}
H. Danino and I. Freund, Phys. Rev. Lett. {\bf 46}, 1127 (1981).
\bibitem{natphys}
K. Tamasaku, Kei Sawada, E. Nishibori and T. Ishikawa, 
Nat. Phys. {\bf 7}, 705-708 (2011).
\bibitem{stanford}
S. Shwartz, R. N. Coffee, J. M. Feldkamp, Y. Feng, J. B. Hastings, 
G. Y. Yin, and S. E. Harris,
Phys. Rev. Lett. {\bf 109}, 013602 (2012).
\bibitem{4a}
K. Tamasaku and T. Ishikawa, Phys. Rev. Lett. {\bf 98}, 244801
(2007).
\bibitem{fano}
U. Fano, Phys. Rev. {\bf 124},1866 (1961).
\bibitem{bianconi}
Alessandra Vittorini-Orgeas, Antonio Bianconi, 
J. Supercond. Nov. Magn. {\bf 22}, 
215 (2009).
\bibitem{4b}
K. Tamasaku, K. Sawada, and T. Ishikawa, 
Phys. Rev. Lett. {\bf 103}, 254801 (2009).
\bibitem{5}
I. Freund and B. F. Levine, Opt. Commun. {\bf 3}, 101 (1971).
\bibitem{henke}
B. L. Henke, E. M. Gullikson, and J. C. Davis, 
{\em X-Ray interactions: photoabsorption, scattering, transmission, 
and reflection at E = 50-30,000 eV, Z = 1-92}, Atomic Data and Nuclear Data Tables, 
vol. 54, no. 2, pp. 181-342, (1993).
\bibitem{nayak}
Maheswar Nayak and Gyanendra S. Lodha, Journal of Atomic, Molecular, and Optical Physics
{\bf 2011}, Article ID 649153 (2011);
http://dx.doi.org/10.1155/2011/649153
\bibitem{6}
B. Barbiellini, J. N. Hancock, C. Monney, 
Y. Joly, G. Ghiringhelli, L. Braicovich, T. Schmitt, 
Phys. Rev. B {\bf 89}, 235138 (2014).
\bibitem{landau}
L. Landau in {\em Niels Bohr and the Development of Physics}, 
edited by W. Pauli (McGraw-Hill, New York, (1955), p. 52.
\bibitem{bogoliubov}
N. N. Bogoliubov and D. V. Shirkov, {\em Quantum Fields}
(Benjamin/Cummings, Reading, MA 1983).
\bibitem{volovik}
G. E. Volovik, {\em The Universe in a Helium Droplet}, (Oxford University Press, 
New York 2008).
\bibitem{jeger} F. Jegerlehner, Nuclear Physics B {\bf 181-182}, 135 (2008).
\bibitem{8}
B. Barbiellini and Piero Nicolini, Phys. Rev. A {\bf 84}, 022509 (2011).
\bibitem{zambianchi}
P. Zambianchi, {\em Nonlinear Response Functions for x-Ray Laser Pulses}, in 
{\em Nonlinear Optics, Quantum Optics and Ultrafast Phenomena with x-Rays: 
Physics with x-Ray Free-Electron Lasers}, 
edited by B.W. Adams (Kluwer Academic Publishers, USA, 2003), p. 287.
\bibitem{joly}
O. Bunau and Y. Joly,
J. Phys. : Condens. Matter {\bf 21}, 345501 (2009).
\bibitem{TDDFT}
O. Bunau and Y. Joly,
Phys. Rev. B {\bf 85}, 155121 (2012).

\bibitem{yabashi01}
M. Yabashi, T. Mochizuki, H. Yamazaki, S. Goto, H. Ohashi, K. Takeshita, 
T. Ohata, T. Matsushita, K. Tamasaku, Y. Tanaka, and T. Ishikawa,
Nucl. Instrum. Methods Phys. Res., Sect. A {\bf 467-468}, 678 (2001).

\bibitem{foot1} An order-of-magnitute estimate obtained from perturbation theory
gives $|\chi^{(2)}_{NR}|=\alpha^2 e r_0 \rho/(k_p k_s k_i^2) = 1.2\times10^{-16}$ cm$^2$/StC.

\bibitem{draine}
B. T. Draine, Astrophys. J. {\bf 598}, 1026 (2003);
[http://www.astro.princeton.edu/\~draine/dust/dust.diel.html]


\bibitem{SM}
See Supplemental Material for rocking curves details.

\bibitem{lodha}
Pragya Tripathi, G.S. Lodha, M.H. Modi, A.K. Sinha, K.J.S. Sawhney, R.V. Nandedkar,
Optics Communications {\bf 211}, 215 (2002).

\bibitem{foot2} Near edge excitation processes may be hindered by the high-intensity regime of XFEL spectroscopy, where multiple ionizations of matter can occur during the XFEL impulse. Therefore, the experimental results should be also validated using the synchrotron radiation regime from storage ring sources.


\end{thebibliography}
\end{document}